\documentstyle[aps,preprint]{revtex}
\begin{document}
\draft
\title{Absolute negative refraction and imaging of unpolarized electromagnetic wave by two-Dimensional photonic crystals}
\author{Xiangdong Zhang}
\address{Department of Physics, Beijing Normal University, Beijing
100875, P. R. China}

\maketitle

\begin{abstract}
Absolute negative refraction  regions for $both$ polarizations of
electromagnetic wave in two-dimensional photonic crystal have been
found through both the analysis and the exact numerical
simulation. Especially, absolute all-angle negative refraction for
$both$ polarizations has also been demonstrated. Thus, the
focusing and image of unpolarized light can be realized by a
microsuperlens consisting of the two-dimensional photonic
crystals. The absorption and compensation for the losses by
introducing optical gain in these systems have also been
discussed.
\end{abstract}
\narrowtext

\section{INTRODUCTION}

Recently there has been a great deal of interest in studying a
novel class of media that has become known as the left-handed
materials[1-19]. These materials are characterized by simultaneous
negative permittivity and permeability. Properties of such
materials were analyzed theoretically by Veselago over 30 years
ago[1], but only recently they were demonstrated
experimentally[4,5]. As was shown by Vesselago, the left-handed
materials possess a number of unusual electromagnetic effects
including negative refraction, inverse Snell's law, reversed
Doppler shift, and reversed Cerenkov radiation. These anomalous
features allow considerable control over light propagation and
open the door for new approaches to a variety of applications.

 It was shown that the negative refraction could also occur in
photonic crystal (PC)[20-32]. In these PC structures, there are
two kinds of cases for negative refraction occuring. The first is
the left-handed behavior as being described above[20-26]. In this
case, $\vec{k}$, $\vec{E}$ and $\vec{H}$ form a left-handed set of
vectors (i.e., $\vec{S}\cdot\vec{k}<0$, where $\vec{S}$ is the
Poynting vectors). Another case is that the negative refraction
can be realized without employing a negative index or a backward
wave effect[27-32]. In this case, the PC is behaving much like a
uniform right-handed medium(i.e. $\vec{S}\cdot\vec{k}>0$). In
particular, Luo $et$ $al.$[27] have shown that all-angle negative
refraction (AANR) could be achieved at the lowest band of
two-dimensional(2D) PC. The advantages of negative refraction in
the lowest valence band are the single-beam propagation and high
transmission efficiency. These can help us to design
microsuperlense and realize the focusing of the wave. Very
recently, the focusing and the image by 2D PC slab have been
demonstrated experimentally [28-30].

It is well known that the electromagnetic (EM) wave can be
decomposed into E polarization (S wave) and H polarization (P
wave) modes for the 2D PC structures [33]. However, the above
discussions about the negative refraction and the focusing of the
wave in the 2D PC all focused on a certain polarized wave, S wave
or P wave. It is a natural question to ask whether or not a
complete negative refraction region for all polarized wave exists?
It is obviously very interesting that the absolute negative
refraction region can be found and the focusing of both polarized
waves be realized in the 2D PC at the same time. In this paper, we
will demonstrate that the complete negative refraction regions for
all polarizations exist in some 2D metallo-dielectric PC
structures. Especially, the absolute AANR has also been found.
Thus, the focusing and image of unpolarized light can be realized
by a microsuperlens consisting of the 2D PC.

The rest of this paper is arranged as follows. In Sec. II, we
demonstrate the absolute all-angle negative refraction behaviors
in 2D PC. The imaging behaviors for both polarized waves are
discussed in Sec. III. A conclusion is given in Sec. IV.

\section{ABSOLUTE NEGATIVE REFRACTION IN 2D PC}

We first consider a 2D square lattice of coated cylinders immersed
in a air background with lattice constant $a$. The coated
cylinders have metallic cores coated with a dielectric coating.
The radii of metallic core and coated cylinder are 0.15a and
0.45a, respectively. The dielectric constant of dielectric coating
is taken as $14$, which can be realized by a mixture of glass
spheres and alumina flake [27]. For the metallic component,
 we use the frequency-dependent dielectric
constant, $\epsilon=1-\frac{f_{p}^2}{f(f+i\gamma)}$. For all
numerical calculations carried out in this work, following Ref.34,
we have chosen $f_{p}=3600$THZ and $\gamma=340$THZ, which
corresponds to a conductivity close to that of Ti. However, our
discussions and conclusions given below can apply to other metal
parameters as well. In order to simplify the problem, we first
consider the cases without absorption ($\gamma=0$). The effect of
absorption will be discussed at the latter part. The band
structures of this system for the S wave and the P wave are
calculated by the multiple-scattering Korringa-Kohn-Rostoker
method. The method has been described in Ref. [35]. The results
are shown in the Fig.1. Solid curves are for the P wave and dotted
curves for the S wave. We focus on the problems of wave
propagation in low frequency band. Owing to the strong scattering
effects, it is generally difficult to describe the propagation
behavior of EM wave in the PC in a simple yet accurate way.
However, a lot of theoretical and experimental
practices[20-22,25,27,31,32] have shown that the overall behavior
of the wave propagation within a PC can be well described by
analyzing the equifrequency surface (EFS) of the band structures,
because the gradient vectors of constant-frequency contours in
k-space give the group velocities of the photonic modes. Thus, the
propagation direction of energy velocity of EM wave can be deduced
from them. The EFS of the above system can also be calculated by
using the multiple-scattering Korringa-Kohn-Rostoker method [35].
The EFS contours  for the S wave and the P wave at several
relevant frequencies are demonstrated in Fig.2.

 It is clear from the figures that the lowest bands
have $\vec{S}\cdot\vec{k}>0$ everywhere within the first Brillouin
zone, meaning that the group velocities ($v_{g}$) are never
opposite to the phase velocity. Some low frequency contours such
as 0.18 and 0.2 are very close to a perfect circle, and the group
velocity at any point of the contour is collimated with the $k$
vector, indicating that the crystal behaves like an effective
homogeneous medium at these long wavelengths. However, the 0.232
contours for both of the S wave and the P wave are significantly
distorted from a circle, which are convex around M points. The
conservation of $k$ component along the surface of refraction
would result in the negative refraction effect in this case. More
interestingly, the EFS contours for the S wave and the P wave at
this frequency are basically same, which leads to the appearance
of negative refraction for both polarized waves at the same time.
It is valuable to design a frequency region in the 2D PC to
realize the negative refraction for both polarized waves. We call
such a region as $absolute$ $negative$ $refraction$ $region$.
 Within this region,
some frequencies (shadow region in Fig.1) are below $0.5\times2\pi
c/a_{s}$ (where $a_{s}$ is the surface-parallel period). According
to the analysis approach of Ref. [27], we find that they satisfy
the required condition for all-angle negative refraction (AANR).
Under these conditions, a EM beam incident on the $\Gamma$M
surface with various incident angle will couple to a single Bloch
mode that propagates into this crystal on the negative side of the
boundary normal. We define a frequency region for the AANR by
using these criteria according to Ref. [27]. The shadow region
(width $2.5\%$ around $\omega=0.232(2\pi c/a)$) in the Fig.1
represents such a common region  for $both$ polarizations of EM
wave (S wave and P wave). We call it $absolute$ $AANR$ $region$.
In this region, a slit beam of S wave and P wave incident on the
$\Gamma$M surface with various incident angle will propagate into
this crystal on the negative refraction side at the same time.

In order to test the above analysis, we do numerical simulations
in the present systems. We take the slab samples which consist of
15-layer coated cylinders in the air background with a square
arrays. The parameters of coated cylinders are the same as the
case in the Fig.1. The shape of the sample and a snapshot of
refraction process are shown on the top of Fig.3. The black frame
marks the size of the sample. The surface of the sample is the
(11) surface. When a slit beam of frequency $\omega=0.232(2\pi
c/a)$ with a half width $3a$ incident normal to the left surface
of the sample, it will be refracted two times by two interfaces of
the slab. There are two kinds of possibility for the refracted
wave. It maybe travel on the path of positive refraction or the
path of negative refraction as shown in figure. The simulations
are based on a highly efficient and accurate multiple-scattering
method [35]. In our simulations, the widths of the samples are
taken enough large, such as 40a, to avoid the edge diffraction
effects. The simulation results for the P wave and the S wave are
plotted in the Fig.3(a) and (b), respectively. The field energy
patterns of incidence and refraction are shown in the figures. The
arrows and texts illustrate the various beam directions. It can be
clearly seen that the energy fluxes of refraction wave outside of
the sample travel on the path of negative refraction. From the
directions of refraction energy flux, we can obtain the refraction
angle and further refraction index. For example, the refraction
angles under the above case at $\omega=0.232(2\pi c/a)$ are about
$35^0$ for both polarized waves, which are consistent with the
estimation from the EFS in the Fig.2. Then, using Snell's law, we
obtain the effective refraction indexes of $-0.872$. Varying the
shape of wedge sample, we have checked the case with different
incident angle. Fig.4 shows the refracted angle $\theta$ versus
incident angle $\theta_0$ at $\omega=0.232(2\pi c/a)$ for the S
wave (circle dots) and the P wave (triangular dots), respectively.
Apart from a small region of zero refracted angle corresponding to
the small incidence angles, all-angle negative refraction can be
observed for both polarized waves at this frequency. Since we know
the optimum frequency for a broad angle negative refraction, we
can use our PC to test the microsuperlens effect.

\section{IMAGING OF UNPOLARIZED ELECTROMAGNETIC WAVE}

  It is well known that an important application of negative refraction
materials is the microsuperlens [3]. Ideally, such a superlens can
focus a point source on one side of the lens into a real point
image on the other side even for the case of a parallel sided slab
of material. However, all discussions about the imaging by 2D PC
microsuperlens had focused on a certain polarized wave, S wave or
P wave. A image of unpolarized light source had not been realized.
Based on the knowledge of the absolute negative refraction, in the
following, we will explore the possibility to realize such a
image.

 In order to model such a superlens, we
take a slab of the sample with 40a width and seven-layer
thickness. A continuous-wave point source is placed at a distance
1.0a from the the left surface of the slab. The frequency of the
incident wave emitting from such a point source is $0.231(2\pi
c/a)$, chosen to lie within the region where absolute all-angle
negative refraction may occur (see Fig.1). We still employ the
multiple-scattering method to calculate the propagation of both
polarized waves in such a system. The typical results of $E_{z}$
field pattern for the S wave and $H_{z}$ field pattern for the P
wave across the slab sample are plotted in Fig.5(a) and (b),
respectively. X and Y present vertical and transverse direction of
wave propagating, respectively. The fields in figures are over
$20a\times 20a$ region around the center of the sample. The
geometries of the PC slab are also displayed. One can find quite a
high quality image formed in the opposite side of the slab. A
closer look at the data reveals a transverse size (full size at
half maximum) of the image spot as $0.9a$ (or $0.21\lambda$) for
the S wave and  $0.85a$ (or $0.2\lambda$) for the P wave. In
particular, the positions of image for both polarized waves are
approximate same. They are about at a distance of $1.0a$ from the
right surface of the slab. That is to say, the image of
unpolarized wave point source can be realized by such a 2D PC
slab.

Our calculations indicate that the high quality images for the
unpolarized wave can be obtained only in some certain frequency
regions, $absolute$ $AANR$ $region$ (shadow region in Fig.1). The
physical reason can be understood by the analysis of Lou $et$
$al.$ [31]. In order to show this feature, we have plotted the
intensity distributions of the S wave and the P wave along the
transverse direction ($y/a$) at the image plane for several
frequencies in Fig.6(a) and (b), respectively. Solid lines in
figures represent the case with $\omega=0.231(2\pi c/a)$)(in the
absolute AANR region). The good focusing effects for both
polarized waves are clearly visible at this case. With increasing
or decreasing the excitation frequencies of the monochromatic
sources, and going beyond the absolute AANR region such as dotted,
dashed  and dot-dashed lines in Fig.6(a) and (b), we find that the
focusing effects degrade gradually and disappear at end.

The above discussion only focused on one kind of
metallo-dielectric system. In fact, a few other systems also
posses the similar character, such as some systems of coated
cylinder with internal metal cylinder coated by semiconductor Ge
($\epsilon=18$) in air background. The common feature of these
systems is that they all include metal component. Therefore, the
absorption for these systems is inevitable. Lou $et$ $al.$ [31]
have pointed that the central image peak disappear and the image
degrade gradually with the increase of absorption. However,
fortunately, the loss can be overcome by introducing the optical
gain in the systems. Recently, Ramakrishna and Pendry [16] have
suggested a method to remove the absorption by introducing optical
gain into the lens made from a multilayers stack of thin
alternating layers of silver and dielectric medium. Here, we
borrow their idea and introduce the optical gain in the 2D PC
superlens. Because it is not sensitive to the absorption and the
gain for the P wave in the present system, we only present the
calculated results for the S wave. Fig.7(a) shows the intensity
distribution as a function of transverse coordinate ($y/a$) for
the S wave at the image plane (1.0a away from the second
interface). Curve {\it A} is corresponded to the case without
absorption and curve {\it B} to that with absorption, in this case
$\gamma$ is taken as $340$THZ. Comparing the curve {\it A} with
the curve {\it B}, we find that the central peak of image decrease
with the introducing of absorption, which is agree with the
analysis of Ref.[31]. The result by introducing gain to remove the
absorption for the corresponding case is plotted in Fig.7(b).
Curves {\it B} in the Fig.7(a) and (b) are the same one, and curve
{\it C} in the Fig.7(b) is the result with the dielectric constant
$\epsilon=14-0.12i$ for the dielectric part of coated cylinder. We
do not find any difference between curve {\it A} in the Fig.7(a)
and curve {\it C} in the Fig.7(b). In fact, for any cases of
absorption, the losses can always be compensated by introducing
fitted gain. Thus, the lens based on the above 2D PC can work well
even in presence of absorption. We would like to point out that
the optical gains being introduced are small and they can not let
the system reach the threshold of lasing. Thus, no lasing
solutions can be produced in present cases. The optical gain here
only plays a role on improving the wave intensity.

\section{CONCLUSION}

Based on the exact numerical simulations and physical analysis, we
have found the absolute negative refraction regions for $both$
polarizations of EM wave in the 2D PC. Especially, the absolute
all-angle negative refraction for $both$ polarizations has also
been demonstrated. The focusing and image of unpolarized light
have been obtained by a microsuperlens consisting of the 2D PC.
Although the losses due to the absorption reduce the intensity of
central peak of the image, it can always be compensated by
introducing fitted optical gain in these systems. Therefore, the
lens based on the 2D PC is applicable even in presence of
absorption. We hope that this work can stimulate interests of
experimental studies of focusing and image of unpolarized light in
the 2D PC.

\begin{center}
Acknowledgments
\end{center}
We wish to thank C. T. Chan for useful discussion. This work was
supported by the National Natural Science Foundation of China
(Grant No.10374009) and the National Key Basic Research Special
Foundation of China under Grant No.2001CB610402. The project
sponsored by SRF for ROCS, SEM and the Grant from Beijing Normal
University.

\newpage

FIGURE CAPTIONS

Fig.1, The calculated photonic band structures of a square lattice
of coated cylinder in air for S wave (dashed lines) and P wave
(solid lines). The radii of the dielectric cylinder with
$\epsilon=14$ and inner metallic cylinder are $R=0.45a$ and
$r=0.15a$, respectively. The light line shifted to M is shown in
dashed line. The shadow represents the AANR region. The inset
shows the microstructure.

Fig.2, Several constant-frequency contours for S wave (solid
lines) and P wave (dotted lines) of the first band of the 2D
coated PC. The structure and parameters of the PC are identical to
those in Fig.1. The numbers in the figure mark the frequencies in
unit of $2\pi c/a$.

Fig.3, Simulation of negative refraction. The shape of the sample
and a snapshot of refraction process are shown on top of the
figure. The intensities of electric field for S wave (a) and
magnetic field for P wave (b) for incidence and refraction are
shown. The 2D PC slabs with 15-layer thickness are marked as dark
dots in figures. The frequencies of incident wave are
$\omega=0.232(2\pi c/a)$ for both polarized waves. The crystal and
parameters are identical to those in Fig.1.

Fig.4, The angles of refraction ($\theta$) versus angles of
incidence ($\theta_0$) at $\omega=0.232(2\pi c/a)$. Circle dots
are corresponded to the S wave and triangular dots to the P wave.
The crystal and parameters are identical to those in Fig.3.

Fig.5, $E_{z}$ field patterns for S wave (a) and $H_{z}$ field
patterns for P wave (b) of point sources and their images across
an seven-layer 2D photonic crystal slab at frequency
$\omega=0.231(2\pi c/a)$. The parameters for cylinders are
identical to those in Fig.3. Dark and bright regions correspond to
negative and positive $E_{z}$ (or $H_{z}$), respectively.

Fig.6, Intensity distribution for S wave (a) and P wave (b) along
the transverse (y) direction at the image plane for several
frequencies shown as insets. The crystal and parameters are
identical to those in Fig.5.

Fig.7, Intensity distribution along the transverse (y) direction
at the image plane. (a) The case with absorption (B) and that
without absorption (A). (a) The case with absorption (B) and that
with absorption and gain (C). The crystal and parameters are
identical to those in Fig.5.

\end{document}